\documentclass[nofootinbib, reprint, preprintnumbers, amsmath, amssymb, amsfonts, aps, pra, superscriptaddress, floatfix]{revtex4-2}
\usepackage[utf8]{inputenc}
\usepackage{mathtools}
\usepackage{graphicx}
\usepackage{dcolumn}
\usepackage{bm}
\usepackage{physics}
\usepackage{float}
\usepackage{placeins}
\usepackage{color}
\usepackage{braket}
\usepackage[normalem]{ulem}
\usepackage{enumerate}
\usepackage{enumitem}
\usepackage[colorlinks=true,linkcolor=blue,citecolor=blue,urlcolor =blue]{hyperref}
\usepackage{mathrsfs}
\usepackage{bbm}
\usepackage{siunitx}
\usepackage{graphicx,color}
\usepackage{amssymb}
\usepackage{amsmath}
\usepackage{placeins} 
\usepackage{subcaption}
\usepackage{bm}
\usepackage{comment}
\usepackage{lipsum}

\begin{document}
\title{
Broadband AC Magnetic Field Sensing via Continuous wave optically detected magnetic resonance with NV Centers in diamond
}

\author{Ryohei Dokai}
\affiliation{Department of Electrical, Electronic, and Communication Engineering, Faculty of Science and Engineering, Chuo university, 1-13-27, Kasuga, Bunkyo-ku, Tokyo 112-8551, Japan}%

\author{Ryusei Okaniwa}
\affiliation{School of Fundamental Science and Technology, Keio University, Yokohama, Kanagawa 223-8522, Japan}
\affiliation{Center for Spintronics Research Network, Keio University, Yokohama, Kanagawa 223-8522, Japan}

\author{Miku Ishizaki}
\affiliation{Department of Electrical, Electronic, and Communication Engineering, Faculty of Science and Engineering, Chuo university, 1-13-27, Kasuga, Bunkyo-ku, Tokyo 112-8551, Japan}

\author{Junko Ishi-Hayase}
\affiliation{School of Fundamental Science and Technology, Keio University, Yokohama, Kanagawa 223-8522, Japan}
\affiliation{Center for Spintronics Research Network, Keio University, Yokohama, Kanagawa 223-8522, Japan}

\author{Yuichiro Matsuzaki}
\affiliation{Department of Electrical, Electronic, and Communication Engineering, Faculty of Science and Engineering, Chuo university, 1-13-27, Kasuga, Bunkyo-ku, Tokyo 112-8551, Japan}%

\date{\today}


\begin{abstract}

The nitrogen--vacancy (NV) center in diamond has attracted considerable attention as a highly sensitive quantum sensor that can operate at room temperature. In particular, continuous-wave optically detected magnetic resonance (CW-ODMR) is promising for a wide range of applications because of its simplicity. However, conventional AC magnetic-field sensing schemes based on CW-ODMR suffer from a limited detection bandwidth: the detectable frequency is either fixed by intrinsic physical parameters of the NV center or, even when tunable, restricted to a narrow range of only a few MHz. 
Here, we propose a broadband AC magnetometry scheme based on CW-ODMR with NV centers using microwave-driven dressed states.
Through theoretical analysis and numerical simulations, we show that the proposed scheme enables the detection of AC magnetic fields with frequencies up to the order of $100~\mathrm{MHz}$, which has been difficult to achieve using conventional CW-ODMR-based methods.

\end{abstract}

\maketitle
\section{Introduction}
\label{sec:intro}
In recent years, quantum sensing -- the application of the principles of quantum mechanics to measurement technology -- has been developing rapidly. Quantum sensing is a technology that exploits the extremely sensitive properties of quantum systems such as atoms and electron spins to measure physical quantities such as magnetic fields, electric fields, and temperature~\cite{Degen2017, Aslam2023}. Compared with classical sensors, this technology is expected to provide improved spatial resolution and sensitivity~\cite{Tanaka2015, Facon2016, Huang2024}.
 
Among the many candidate quantum sensors, the nitrogen--vacancy (NV) center in diamond has attracted attention as a promising solid-state quantum sensor~\cite{Barry2020}. An NV center is a complex defect in which one carbon atom in the diamond lattice is replaced by a nitrogen atom (N) and an adjacent lattice site is a vacancy (V). The electron spin trapped at the NV center can maintain quantum coherence (a superposition of quantum states) for a long time at room temperature and atmospheric pressure~\cite{Balasubramanian2009, Herbschleb2019}, and the spin state can be initialized to a particular state ($m_s=0$) by irradiation with green laser light~\cite{Harrison2004}. The difference in fluorescence intensity between spin states enables optical readout~\cite{Gruber1997, Schirhagl2014}.
 
Furthermore, the spin state of the NV center can be manipulated using microwaves (MW)~\cite{Schirhagl2014}, and many practical magnetic-field sensing schemes using NV centers have been proposed and demonstrated, including wide-field imaging~\cite{Pham2011, LeSage2013, Mizuno2020}, AFM (atomic force microscope)-based methods~\cite{Balasubramanian2008, Degen2008, Grinolds2011, Huxter2022}, and vector magnetic-field sensing~\cite{Pham2011, Maertz2010, Steinert2010, Kitazawa2017, Yahata2019, Wang2021}.
 
Methods for AC magnetic-field sensing using NV centers can be broadly classified into pulse methods and continuous-wave (CW) methods. Pulse methods manipulate quantum states using pulse sequences and can achieve high sensitivity, but 
they require complex preliminary calibration such as strength and phase adjustment of microwaves for controlling the rotation speed and axis in Bloch-sphere~\cite{Balasubramanian2008, Taylor2008, Maze2008, PhamDD2012, Loretz2013, Wolf2015, Stark2017}. In contrast, the CW-ODMR method is a simple technique in which the fluorescence intensity is measured while continuously irradiating the sample with both laser light and microwaves. Previous studies have demonstrated successful detection of AC (time-varying) magnetic fields in the MHz band using CW-ODMR~\cite{Saijo2018, Yamaguchi2019}. However, this scheme has the drawback that the detectable frequency is fixed by the physical constants such as intrinsic strain and bias magnetic field strength of the NV center. To address this, a method has been proposed and demonstrated in which the detection frequency is made tunable by applying a radio frequency (RF) field, but the tunable range was limited to a few MHz~\cite{Okaniwa2024}.
 
Here, we propose a new sensing scheme based on CW-ODMR that overcomes the bandwidth limitation. Specifically, we use microwave-driven dressed states and vary the bandwidth by adjusting the Rabi frequency. Since microwave driving allows the Rabi frequency to be increased to about $100~\mathrm{MHz}$ without violating the rotating-wave approximation~\cite{Jung2025}, the proposed scheme makes it possible to detect AC magnetic fields up to the order of $100~\mathrm{MHz}$, which has been difficult to achieve previously. We quantitatively evaluate its performance through numerical calculations.

\section{The Diamond NV Center}

In this chapter, we describe the physical model of the diamond NV center treated in this work, the basic equations describing its dynamics, and the theory of the noise that limits the measurement sensitivity.

\subsection{Hamiltonian and Eigenstates of the NV Center}

The electron spin of the diamond NV center is a system with spin quantum number $S=1$, and its energy states are described by a Hamiltonian. In the presence of an external magnetic field $\bm{B}$, the NV center Hamiltonian $\hat{H}$ is expressed in terms of the zero-field-splitting term and the Zeeman term as~\cite{Barry2020}
\begin{equation}
\hat{H} = D\hat{S}_z^{2} + E(\hat{S}_x^{2} - \hat{S}_y^{2}) + \gamma_e \bm{B}\cdot\hat{\bm{S}},
\label{eq:nv_static}
\end{equation}
where $D$ is the zero-field-splitting parameter ($D/2\pi \approx 2.87$~GHz), $E$ is the crystal-strain parameter ($E/2\pi \approx 10$~MHz), $\gamma_e$ is the electron gyromagnetic ratio ($\gamma_e/2\pi \approx 28$~GHz/T), and $\hat{\mathbf{S}} = (\hat{S}_x, \hat{S}_y, \hat{S}_z)$ are the spin operators.
\begin{equation}
\begin{split}
\hat{S}_x &= \ket{B}\bra{0} + \ket{0}\bra{B}, \quad
\hat{S}_y = -i\ket{D}\bra{0} + i\ket{0}\bra{D}, \\
\hat{S}_z &= \ket{B}\bra{D} + \ket{D}\bra{B}.
\end{split}
\label{eq:S_def}
\end{equation}
Here we have defined the bright state $\ket{B}$ and the dark state $\ket{D}$ as the following linear combinations of the states $\ket{+1}$ and $\ket{-1}$ in the $\hat{S}_z$ basis:
\begin{equation}
\ket{B} = \frac{1}{\sqrt{2}}\bigl(\ket{+1} + \ket{-1}\bigr), \quad
\ket{D} = \frac{1}{\sqrt{2}}\bigl(\ket{+1} - \ket{-1}\bigr).
\label{eq:bright_dark}
\end{equation}

\subsection{Schr\"{o}dinger Equation and Master Equation}

The time evolution of a quantum-system state is described by the Schr\"{o}dinger equation in the case of a closed system. In reality, however, the NV center is an open quantum system that constantly interacts with its surrounding environment. In such systems, interaction with the external environment disturbs the phase coherence of the quantum state and dissipates energy. 
To describe these effects, we use the Lindblad master equation, which extends the Schr\"{o}dinger equation~\cite{Lindblad1976,Gorini1976,Hornberger2009}:
\begin{equation}
\frac{d\rho}{dt} = -\frac{i}{\hbar}[\hat{H}, \rho] + \sum_k \mathcal{L}_k(\rho).
\label{eq:lindblad}
\end{equation}
Here $\rho$ is the density operator and $\mathcal{L}_k(\rho)$ is the Lindblad term, expressed as
\begin{equation}
\mathcal{L}_k(\rho) = \Gamma_k\!\left(\hat{L}_k \rho \hat{L}_k^{\dagger} - \tfrac{1}{2}\{\hat{L}_k^{\dagger}\hat{L}_k, \rho\}\right),
\label{eq:lindblad_kernel}
\end{equation}
where $\hat{L}_k$ is the jump operator corresponding to a relaxation process and $\Gamma_k$ is the corresponding relaxation rate. Throughout this paper, we set $\hbar = 1$ and express all Hamiltonians in angular-frequency units. By using this equation, it becomes possible to numerically simulate the time evolution in a realistic system that includes initialization by laser irradiation and relaxation processes. By solving this equation and computing the steady-state density matrix, the occupation probability $P$ of the ground state $\ket{0}$ can be obtained.

\subsection{Sensing Methods Using NV Centers and Measurement Noise}

Here we outline measurement noise~\cite{Taylor2008, Kitazawa2017}. Optical readout of the spin state is not perfect, owing to the limit of collection efficiency and the nature of spontaneous emission~\cite{Barry2020}. The information on the NV center is transferred to photons, and the photon state is described as~\cite{Taylor2008, Kitazawa2017}
\begin{align}
\rho_0^{(ph)} &= (1-\tilde{\alpha}_0)\ket{0}_{ph}\bra{0} + \tilde{\alpha}_0 \ket{1}_{ph}\bra{1}, \label{eq:rho_ph0}\\
\rho_1^{(ph)} &= (1-\tilde{\alpha}_1)\ket{0}_{ph}\bra{0} + \tilde{\alpha}_1 \ket{1}_{ph}\bra{1}, \label{eq:rho_ph1}
\end{align}
where $\rho_0^{(ph)}$ ($\rho_1^{(ph)}$) is the state of the photon emitted from the NV center when the NV state is $\ket{0}$ ($\ket{B}$ or $\ket{D}$), and $\tilde{\alpha}_0$ ($\tilde{\alpha}_1$) is the probability that a photon is emitted when the NV state is $\ket{0}$ ($\ket{B}$ or $\ket{D}$). Also, $\ket{0}_{ph}$ and $\ket{+1}_{ph}$ denote the Fock states of the photon. Strictly, there are cases in which two or more photons are emitted, but we ignore such higher-order terms by assuming large photon loss. In this work, we adopt the typical values $\tilde{\alpha}_0 \simeq 0.03620$ and $\tilde{\alpha}_1 \simeq 0.01949$~\cite{Taylor2008}. Then, the expected photon number $\langle \hat{N} \rangle$ measured is
\begin{equation}
\langle \hat{N} \rangle \simeq P\tilde{\alpha}_0 + (1-P)\tilde{\alpha}_1,
\label{eq:N_expectation}
\end{equation}
where $P$ is the probability that the NV center is in the state $\ket{0}$.
 
Now we consider the measurement error. Suppose first that the readout of the spin state is ideal and that we can perfectly distinguish whether the final state is $\ket{0}$ (with probability $P$) or otherwise (with probability $1-P$). The statistical fluctuation (standard deviation $\Delta P$) of the probability $P$ obtained by measurement is then given by the square root of the variance,
\begin{equation}
\Delta P = \sqrt{\frac{P(1-P)}{N_m}},
\label{eq:DeltaP}
\end{equation}
where $N_m$ is the number of measurements. Between the small change $\Delta P$ in $P$ and the magnetic-field measurement error $\delta B$, the relation
\begin{equation}
\Delta P = \left|\frac{dP}{dB}\right|\delta B
\label{eq:DeltaP_dB}
\end{equation}
holds. Solving this for $\delta B$ and substituting $\Delta P$ from \eqref{eq:DeltaP}, we obtain
\begin{equation}
\delta B = \frac{\sqrt{P(1-P)}}{\left|\dfrac{dP}{dB}\right|\sqrt{N_m}}.
\label{eq:dB_ideal}
\end{equation}
This is the formula for the measurement error in the absence of measurement noise.
 
On the other hand, in the presence of measurement noise due to imperfect optical readout, the measurement error is calculated as
\begin{equation}
\delta B = \frac{\sqrt{\langle\delta \hat{N} \delta \hat{N}\rangle}}{\left|\dfrac{\partial \langle \hat{N}\rangle}{\partial B}\right|\sqrt{N_m}},
\label{eq:dB_noise_general}
\end{equation}
where $\langle\delta \hat{N} \delta \hat{N}\rangle$ is the variance of the photon number, $\langle \hat{N} \rangle$ is the expected photon number, and the number operator $\hat{N}$ can be written as
\begin{equation}
\hat{N} = \sum_{n=0}^{\infty} n \ket{n}_{ph}\bra{n} \simeq \sum_{n=0}^{1} n \ket{n}_{ph}\bra{n} = \ket{1}_{ph}\bra{1},
\label{eq:N_op}
\end{equation}
where $\ket{n}_{ph}$ is the Fock state of the photon, and we have dropped higher-order terms by assuming that the number of photons under consideration is sufficiently small. Therefore, we can take $\langle\delta \hat{N} \delta \hat{N}\rangle \simeq \langle \hat{N} \rangle$, and the measurement error finally becomes
\begin{equation}
\delta B = \frac{\sqrt{\langle \hat{N} \rangle}}{\left|\dfrac{\partial \langle \hat{N}\rangle}{\partial B}\right|\sqrt{N_m}}.
\label{eq:dB_noise}
\end{equation}

\section{CW-ODMR Sensing Schemes}

This chapter reviews in detail the principle of the sensing techniques based on continuous-wave optically detected magnetic resonance (CW-ODMR), which forms the foundation of the present work, and the main schemes proposed in prior studies. We describe the physical mechanisms of previously proposed dressed-state methods and the frequency-tunable scheme, and discuss the bandwidth limitations of the current state of the art.

\subsection{The Measurement Principle of CW-ODMR}

As mentioned earlier, an NV center in diamond is a complex defect consisting of a substitutional nitrogen atom (N) and an adjacent vacancy (V)\cite{Doherty2013}. The electron spin state of the NV center can be both initialized and read out optically\cite{Barry2020}. The magnetic-resonance technique that exploits this property is called optically detected magnetic resonance (ODMR).
 
The ground state of the NV center is a spin triplet ($S=1$) with three sublevels of spin quantum number $m_s=0,\pm1$. Even in the absence of an external magnetic field and strain, the zero-field splitting ($D/2\pi\approx 2.87~\mathrm{GHz}$) produces an energy difference between the $\ket{0}$ state and the $\ket{B}$ ($\ket{D}$) states. When green laser light (around $532~\mathrm{nm}$ wavelength) is applied, the NV center is excited to an optical excited state, after which the spin is preferentially transferred (initialized) into the $m_s=0$ state. Moreover, the fluorescence from the $\ket{0}$ state is brighter than that from the $\ket{B}$ ($\ket{D}$) states.
 
In CW-ODMR, the laser is applied continuously to keep the spin initialized in $\ket{0}$, while microwaves are continuously swept and applied. When the microwave frequency matches (resonates with) the energy difference between spin sublevels, the spin is driven to the $\ket{B}$ ($\ket{D}$) state. As a result, the population of the bright $m_s=0$ state decreases and the observed fluorescence intensity drops. By detecting this fluorescence drop (a \textit{dip}) as a spectrum, the resonant frequency of the spin can be identified -- this is the basic principle of CW-ODMR. In the presence of an external magnetic field, the resonance frequencies shift due to the Zeeman effect, so the magnetic-field strength can be inferred from the displacement of the dips.

\subsection{Sensing Using Dressed States}

Conventional CW-ODMR detects DC and low-frequency magnetic fields by reading the shift of the microwave resonant frequency\cite{Taylor2008,Acosta2009}, but a different approach is needed to detect high-frequency AC magnetic fields in the MHz band. A previous study demonstrated an AC magnetic-field sensing scheme that exploits transitions between sublevels in the NV ground state~\cite{Saijo2018, Yamaguchi2020, Mikawa2023}.
 
The NV center has crystal strain that lifts the degeneracy of $m_s=\pm1$, and forms new eigenstates: the bright state $\ket{B}$ and the dark state $\ket{D}$. In the absence of a DC magnetic field, the energy difference between these levels is given by $2E$ (where $E$ is the strain parameter). 
When an AC magnetic field with frequency $\omega_{\rm AC}$ equal to this energy difference $2E$, polarized along the N-V axis, is applied to such a system, Rabi oscillations are induced between $\ket{B}$ and $\ket{D}$.
Analysis of the Hamiltonian under the simultaneous application of this AC magnetic field and the microwave shows that the AC magnetic field causes splitting of the ODMR spectrum, as previously demonstrated~\cite{Saijo2018}. Specifically, when the AC magnetic field is present, the resonance condition for the dips in the ODMR spectrum becomes
\begin{equation}
\omega_{MW} \simeq D \pm E \pm \tfrac{1}{2}\gamma_e B_{AC},
\label{eq:dressed_resonance}
\end{equation}
where $\gamma_e$ is the electron gyromagnetic ratio and $B_{AC}$ is the amplitude of the AC magnetic field. The advantage of this scheme is its simplicity: AC magnetic fields in the MHz band can be detected using only continuous microwave and laser irradiation, without requiring complex pulse-sequence control. 
Experimentally, a sensitivity of $2.5~\mu\mathrm{T}/\sqrt{\mathrm{Hz}}$ 
has been achieved for an AC magnetic field with $\omega_{AC}/2\pi \approx 4~\mathrm{MHz}$~\cite{Saijo2018}.
 
However, the detectable frequency in this scheme is uniquely determined by the strain parameter $E$ of the NV center. Detection requires satisfying the resonance condition $\omega_{AC}\approx 2E$, and $E$ is a fixed value that depends on the individual diamond sample and the local environment of the NV center.
The values of $E$ can be as large as $E/2\pi =$10MHz~\cite{pettit2017coherent}.
Although there is a method for tuning the frequency by applying an electric field to vary $E$~\cite{Dolde2011}, the device structure becomes more complex due to the addition of electrodes, which may compromise the simplicity advantage of the technique.

\subsection{Frequency-Tunable Scheme Based on Doubly Dressed States}

To overcome the fixed-frequency limitation of the above scheme, a method has been proposed in which an additional control RF magnetic field is used to tune the detection frequency~\cite{Okaniwa2024}.
 
In this method, a control RF magnetic field resonant with the $\ket{B}\leftrightarrow\ket{D}$ transition (frequency $\omega_{AC}\approx 2E$) is first applied. The coherent interaction of the spin with this strong RF field generates new energy eigenstates known as \textit{RF-dressed states}. The energy splitting $\Omega$ between the RF-dressed states depends on the amplitude $B_{RFc}$ of the applied control field,
\begin{equation}
\Omega \simeq 2E \pm \gamma_e B_{RFc}.
\label{eq:rf_dressed_split}
\end{equation}
A target AC magnetic field to be detected is then additionally applied. If the target frequency $\omega_{AC}$ matches this splitting $\Omega$, a further resonance transition is induced between the RF-dressed states, forming a \textit{doubly dressed state}.
 
The feature of this scheme is that the splitting $\Omega$ of the RF-dressed states can be controlled at will by adjusting the control-field amplitude $B_{RFc}$. In other words, by setting $B_{RFc}$ such that $\Omega\approx \omega_{AC}$ for the desired target frequency, magnetic fields at that frequency can be detected. Experimentally, sweeping the control-field strength as a function of the target frequency reveals splitting of the dips in the ODMR spectrum, demonstrating operation as a frequency-tunable AC magnetic-field sensor.

\subsection{Bandwidth Limitations of Existing Schemes}

Although the previous study made frequency-tunable sensing possible in principle, practical bandwidth limitations remain. According to the report, the bandwidth over which this scheme maintains a useful sensitivity is about $7.5~\mathrm{MHz}$~\cite{Okaniwa2024}. The main cause of this bandwidth limitation is believed to be the breakdown of the rotating-wave approximation. To detect a higher-frequency target field, one must create a correspondingly larger energy splitting, which requires a stronger control field $B_{RFc}$. However, when the field strength becomes non-negligible compared with the resonant frequency, the rotating-wave approximation (RWA) on which the theory is based begins to break down. Once the RWA fails, higher-order effects appear: the simple resonance condition is no longer satisfied, and unintended transitions are induced, reducing sensitivity. For this reason, the previous study suggests a practical upper limit of about $13~\mathrm{MHz}$~\cite{Okaniwa2024}.
 
In this way, the approach of simply increasing the control-field strength to widen the splitting of dressed states has fundamental limits. Therefore, in order to detect AC magnetic fields on the order of $100~\mathrm{MHz}$, it is essential to introduce a new physical approach that does not rely on approach that circumvent the limitations imposed by the RWA.

\section{Theory of a Broadband AC Magnetic Field Sensor via CW-ODMR with Microwave-Driven Dressed States}\label{ch:theory}

Here, we present our proposed scheme: a broadband AC magnetic-field sensor based on CW-ODMR using microwave-driven dressed states.
 
The system considered here consists of the electron spin ($S=1$) of a diamond NV center subjected to microwave driving and a target AC magnetic field. The Hamiltonian is defined as
\begin{equation}
\hat{H} = D\hat{S}_z^{2} + E(\hat{S}_x^{2}-\hat{S}_y^{2}) + \lambda_D \hat{S}_y \cos\omega_D t + \lambda_T \hat{S}_z \cos\omega_T t.
\label{eq:H_full}
\end{equation}
The meaning of each term is as follows: $D$ is the zero-field splitting of the NV center ($D/2\pi \approx 2870~\mathrm{MHz}$), $E$ is the strain parameter of the NV center ($E/2\pi \approx 10~\mathrm{MHz}$), $\lambda_D$ is the Rabi frequency (driving strength) of the control microwave, $\omega_D$ is the angular frequency of the control microwave, and $\lambda_T$, $\omega_T$ are the strength and angular frequency of the target AC magnetic field to be detected. All parameters $D$, $E$, $\lambda_D$, $\lambda_T$, $\omega_D$, and $\omega_T$ are defined as angular frequencies, following the convention used throughout this paper.
 
To analyze the Hamiltonian containing time-dependent terms, we transform to an appropriate rotating frame. We first introduce the unitary transformation $U = e^{i\omega_D t \ket{D}\bra{D}}$ into the rotating frame associated with the $\ket{D}$ level. After this transformation and the application of the rotating-wave approximation (RWA), the Hamiltonian is approximated as follows. 
The condition for the RWA to be valid is $\lambda_D \ll \omega_D$. 
Since $\omega_D/2\pi$ is on the order of $2.88~\mathrm{GHz}$, the RWA 
is expected to remain valid even when $\lambda_D/2\pi$ is on the order 
of $100~\mathrm{MHz}$~\cite{Jung2025}.
\begin{align}
\hat{H} &\simeq (D+E)\ket{B}\bra{B} + (D-E-\omega_D)\ket{D}\bra{D} + \frac{\lambda_D}{2}\hat{S}_y \nonumber\\
&\quad + \lambda_T \bigl(\ket{B}\bra{D}e^{-i\omega_D t} + \ket{D}\bra{B}e^{i\omega_D t}\bigr)\cos\omega_T t.
\label{eq:H_rotating}
\end{align}
To simplify the system, we set the microwave frequency $\omega_D$ resonant with the energy of the $\ket{D}$ level, i.e.\ $\omega_D = D-E$. Under this condition the Hamiltonian reduces to
\begin{align}
\hat{H} &\simeq (D+E)\ket{B}\bra{B} + \frac{\lambda_D}{2}\hat{S}_y \nonumber\\
&\quad + \lambda_T\bigl(\ket{B}\bra{D}e^{-i\omega_D t} + \ket{D}\bra{B}e^{i\omega_D t}\bigr)\cos\omega_T t.
\label{eq:H_rotating_resonant}
\end{align}
Further manipulation yields
\begin{align}
\hat{H} &\simeq (D+E)\ket{B}\bra{B} + \frac{\lambda_D}{2}\hat{S}_y \nonumber\\
&\quad + \frac{\lambda_T}{2}\ket{B}\bigl(\bra{D} + i\bra{0}\bigr)e^{-i\omega_D t}\cos\omega_T t + h.c. \nonumber\\
&\quad + \frac{\lambda_T}{2}\ket{B}\bigl(\bra{D} - i\bra{0}\bigr)e^{-i\omega_D t}\cos\omega_T t + h.c.,
\label{eq:H_expanded}
\end{align}
where $h.c.$ denotes the Hermitian conjugate.
 
Next we move to the interaction picture defined by the time-independent Hamiltonian $\hat{H}_0 = (D+E)\ket{B}\bra{B} + \tfrac{\lambda_D}{2}\hat{S}_y$. Setting
\begin{align}
\hat{H}' &= \frac{\lambda_T}{2}\ket{B}\bigl(\bra{D} + i\bra{0}\bigr)e^{-i\omega_D t}\cos\omega_T t + h.c. \nonumber\\
&\quad + \frac{\lambda_T}{2}\ket{B}\bigl(\bra{D} - i\bra{0}\bigr)e^{-i\omega_D t}\cos\omega_T t + h.c.,
\end{align}
and computing the interaction-picture Hamiltonian $\hat{H}_I(t) = e^{i\hat{H}_0 t}\hat{H}'(t)e^{-i\hat{H}_0 t}$, we obtain
\begin{align}
\hat{H}_I(t) &= \frac{\lambda_T}{2}e^{-i\omega_D t + i(D+E)t + i\frac{\lambda_D}{2}t}\cos\omega_T t \nonumber\\
&\quad \times \ket{B}(\bra{D}+i\bra{0}) + h.c. \nonumber\\
&\quad + \frac{\lambda_T}{2}e^{-i\omega_D t + i(D+E)t - i\frac{\lambda_D}{2}t}\cos\omega_T t \nonumber\\
&\quad \times \ket{B}(\bra{D}-i\bra{0}) + h.c.
\label{eq:H_I}
\end{align}
Here,
\begin{align}
&\frac{\lambda_T}{2} e^{-i\omega_D t + i(D+E)t \mp i\frac{\lambda_D}{2}t}\cos\omega_T t \, \ket{B}\bigl(\bra{D} \mp i\bra{0}\bigr) \nonumber\\
&= \frac{\lambda_T}{4} e^{-i\omega_D t + i(D+E)t \mp i\frac{\lambda_D}{2}t}\bigl(e^{i\omega_T t} + e^{-i\omega_T t}\bigr) \nonumber\\
&\quad \times \ket{B}\bigl(\bra{D} \mp i\bra{0}\bigr) \nonumber\\
&= \frac{\lambda_T}{4}\bigl(e^{i(\omega_T - \omega_D + D + E \mp \frac{\lambda_D}{2})t} + e^{-i(\omega_T + \omega_D - D - E \pm \frac{\lambda_D}{2})t}\bigr) \nonumber\\
&\quad \times \ket{B}\bigl(\bra{D} \mp i\bra{0}\bigr) \nonumber\\
&= \frac{\lambda_T}{4}\bigl(e^{i(\omega_T + 2E \mp \frac{\lambda_D}{2})t} + e^{-i(\omega_T - 2E \pm \frac{\lambda_D}{2})t}\bigr)\ket{B}\bigl(\bra{D} \mp i\bra{0}\bigr).
\label{eq:expand_full}
\end{align}
Applying Fermi's golden rule to this expression, the following resonance condition is derived:
\begin{equation}
\omega_T = \left|2E \pm \frac{\lambda_D}{2}\right|
\label{eq:resonance_condition}
\end{equation}
This result indicates that a resonance signal is detected when the target frequency $\omega_T$ matches a frequency determined by the combination of the strain-induced splitting $2E$ of the NV center and the control-microwave strength $\lambda_D$. 
In other words, by tuning the strength parameter $\lambda_D$ of the control 
field, the detectable resonant frequency can in principle be shifted. 
Furthermore, $\lambda_D/2\pi$ can be increased up to about $100~\mathrm{MHz}$ 
without breaking the rotating-wave approximation~\cite{Jung2025}.
Therefore, the detectable frequency band can also be extended up to the order of $100~\mathrm{MHz}$.

\section{Numerical Simulation and Results}\label{ch:simulation}
Here, to verify the validity of our proposal, 
we performed numerical simulations using QuTiP\cite{qutip1}, a Python library for the dynamics of open quantum systems. 
We describe the results in detail and provide a physical interpretation of the obtained data.

\subsection{Simulation Conditions and Parameter Settings}

Following the angular-frequency convention introduced in Chapter~\ref{ch:theory}, 
we additionally include the relaxation rate $\Gamma$ in this category. 
The numerical values of all parameters in the text, figures, and figure 
captions are presented divided by $2\pi$ in units of MHz, and the target 
frequency on the horizontal axis of the spectra is denoted by 
$f_T \equiv \omega_T/2\pi$.

To accurately compute the time evolution of the NV-center quantum state, we used the following Hamiltonian without invoking the rotating-wave approximation (RWA):
\begin{align}
\hat{H} &= (D+E)\ket{B}\bra{B} + \frac{\lambda_D}{2}\hat{S}_y \nonumber\\
&\quad + \frac{\lambda_D}{2}\bigl(-ie^{2i\omega_D t}\ket{D}\bra{0} + ie^{-2i\omega_D t}\ket{0}\bra{D}\bigr) \nonumber\\
&\quad + \lambda_T \bigl(\ket{B}\bra{D}e^{-i\omega_D t} + \ket{D}\bra{B}e^{i\omega_D t}\bigr)\cos\omega_T t.
\label{eq:H_sim}
\end{align}
In Eq.~\eqref{eq:H_sim}, the third term represents the counter-rotating contribution that is dropped under the RWA in the analytical treatment of Chapter~\ref{ch:theory}, but is retained here so that we can quantitatively verify the validity of the RWA used in the theoretical derivation.

Using Eq.~\eqref{eq:H_sim}, we numerically solved the following Lindblad master equation:
\begin{equation}
\frac{d\rho}{dt} = -\frac{i}{\hbar}[\hat{H}, \rho] + \Gamma\sum_{k=1,2}\!\left(\hat{L}_k \rho \hat{L}_k^{\dagger} - \tfrac{1}{2}[\hat{L}_k^{\dagger}\hat{L}_k, \rho]\right).
\label{eq:lindblad_sim}
\end{equation}
For the jump operators $\hat{L}$, we set the following two operators, accounting for relaxation from the bright and dark states ($\ket{B}$, $\ket{D}$) to the ground state ($\ket{0}$):
\begin{equation}
\hat{L}_1 = \ket{0}\bra{B}, \qquad \hat{L}_2 = \ket{0}\bra{D}.
\label{eq:jumps}
\end{equation}
These represent stochastic transitions from $\ket{B}$ and $\ket{D}$, respectively, to the state $\ket{0}$, corresponding to the optical-pumping process.

The flow of the calculation is as follows: we take the initial state to be the $m_s=0$ state ($\ket{0}$), and compute the time evolution up to time $t$ to obtain the time-averaged occupation probability of $\ket{0}$. Specifically, denoting the density matrix at time $t$ by $\rho(t)$, we plot
\begin{equation}
P_0 = \frac{1}{t_{av}}\int_{t_s}^{t_s + t_{av}} \mathrm{Tr}\bigl[\ket{0}\bra{0}\rho(t')\bigr]\,dt'.
\label{eq:P0}
\end{equation}
Then, we vary the target frequency $f_T$ and consider the final occupation probability of $\ket{0}$ at each frequency to obtain the ODMR spectrum. We set $t_{av} = 1.0~\mu\mathrm{s}$ and $t_s = 4.0~\mu\mathrm{s}$, so that the total evolution time is $t = t_s + t_{av} = 5.0~\mu\mathrm{s}$, and use this $P_0$ when computing the sensitivity numerically.

\begin{figure}[ht]
\centering
\includegraphics[width=0.95\linewidth]{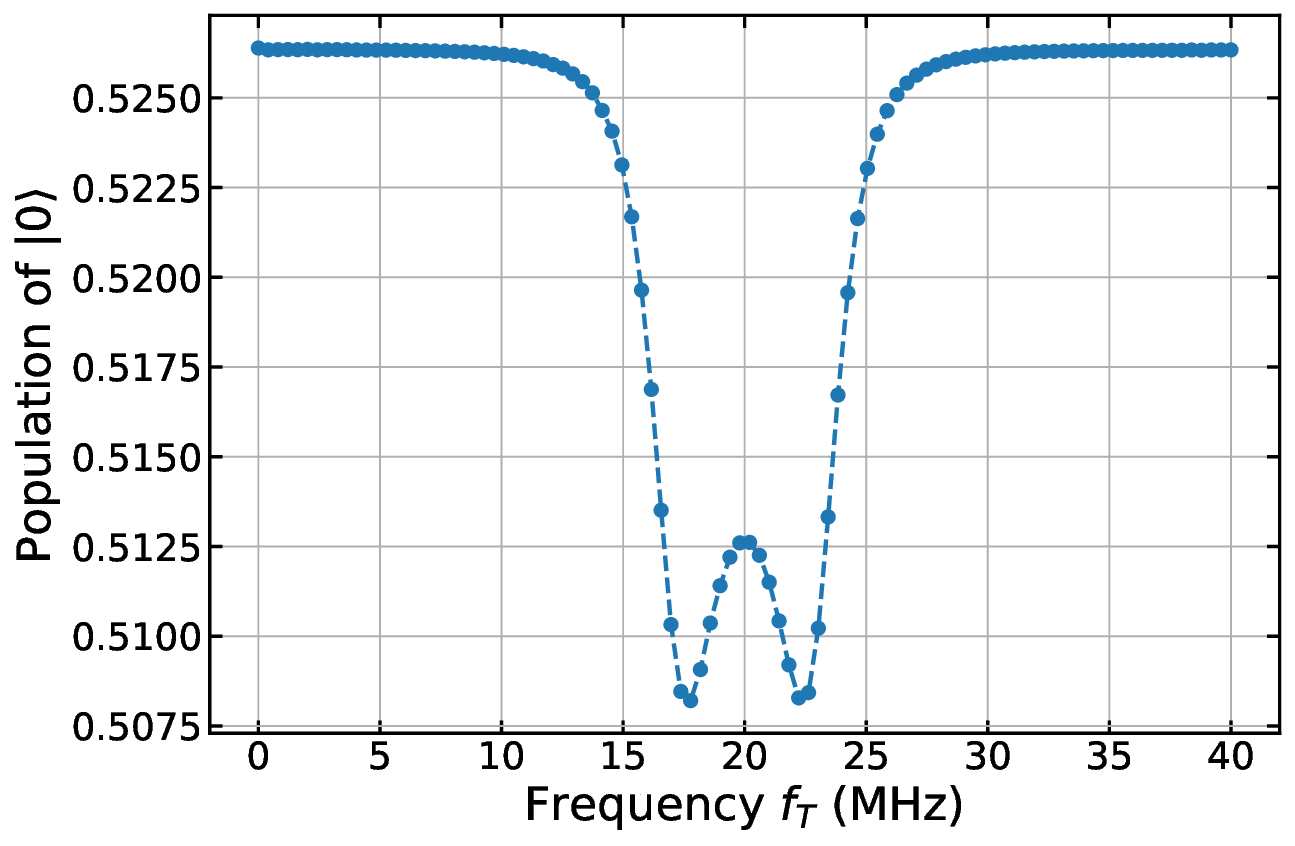}
\caption{Plot of the variation of the $\ket{0}$ ground-state occupation as a function of target frequency $f_T \equiv \omega_T/2\pi$. The parameters are $D/2\pi=2870~\mathrm{MHz}$, $E/2\pi=10~\mathrm{MHz}$, $\omega_D/2\pi=2860~\mathrm{MHz}$, $\lambda_D/2\pi=6.0~\mathrm{MHz}$, $\lambda_T/2\pi=1.0~\mathrm{MHz}$, $\Gamma/2\pi=2.0~\mathrm{MHz}$, and $t=5~\mu\mathrm{s}$.}
\label{fig:odmr_spectrum}
\end{figure}
Figure~\ref{fig:odmr_spectrum} shows the result of the simulation performed with the above parameters. From the figure, the occupation probability drops at specific positions of the target frequency $f_T$. 
Substituting the present simulation parameters ($2E/2\pi=20~\mathrm{MHz}$, $\lambda_D/4\pi=3~\mathrm{MHz}$) into the resonance condition $\omega_T = |2E\pm \lambda_D/2|$ derived in Chapter~\ref{ch:theory}, the predicted resonance frequencies are
\begin{equation}
f_T \approx 
17~\mathrm{MHz},\; 23~\mathrm{MHz}.
\label{eq:predicted_freqs}
\end{equation}
In the simulation results as well, clear dips are confirmed near $17~\mathrm{MHz}$ and $23~\mathrm{MHz}$, in agreement with this prediction.
 
\begin{figure}[ht]
\centering
\includegraphics[width=0.95\linewidth]{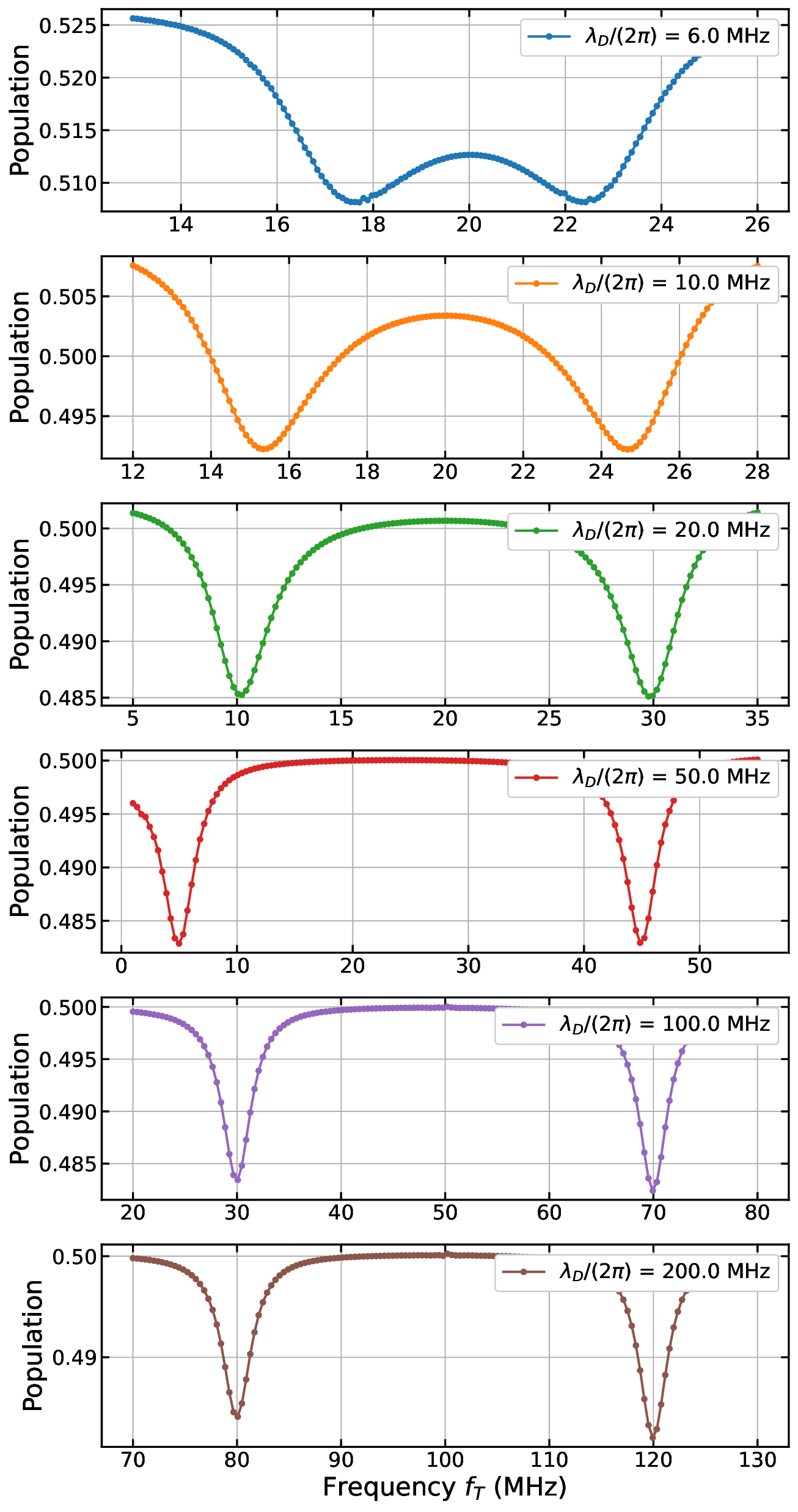}
\caption{Plots of the variation of the $\ket{0}$ ground-state occupation as a function of target frequency $f_T$ when the control-microwave strength $\lambda_D/2\pi$ is varied from $6.0~\mathrm{MHz}$ to $200.0~\mathrm{MHz}$. The other parameters are $D/2\pi=2870~\mathrm{MHz}$, $E/2\pi=10~\mathrm{MHz}$, $\omega_D/2\pi=2860~\mathrm{MHz}$, $\lambda_T/2\pi=1.0~\mathrm{MHz}$, $\Gamma/2\pi=2.0~\mathrm{MHz}$, and $t=5~\mu\mathrm{s}$.}
\label{fig:lambdaD_sweep}
\end{figure}

Furthermore, to verify the broadening of the detectable frequency, which is the key feature of the proposed scheme, Figure~\ref{fig:lambdaD_sweep} shows the simulation results when the control-microwave strength $\lambda_D/2\pi$ is varied from $6.0~\mathrm{MHz}$ to $200.0~\mathrm{MHz}$. The figure shows that as the value of $\lambda_D$ increases, the spacing between the two dips broadens and the resonance frequencies shift to the higher-frequency side. In all cases, the dip positions agree with the resonance condition $\omega_T = \pm(2E\pm \lambda_D/2)$, just as for $\lambda_D/2\pi = 6.0~\mathrm{MHz}$. This result supports that, by using strong microwave driving, we can overcome the conventional bandwidth limit of a few MHz and that broadband AC magnetic-field sensing up to about $100~\mathrm{MHz}$ is possible in principle.
 
\subsection{Amplitude Dependence of the Signal and Fermi's Golden Rule}

\begin{figure}[ht]
\centering
\includegraphics[width=0.95\linewidth]{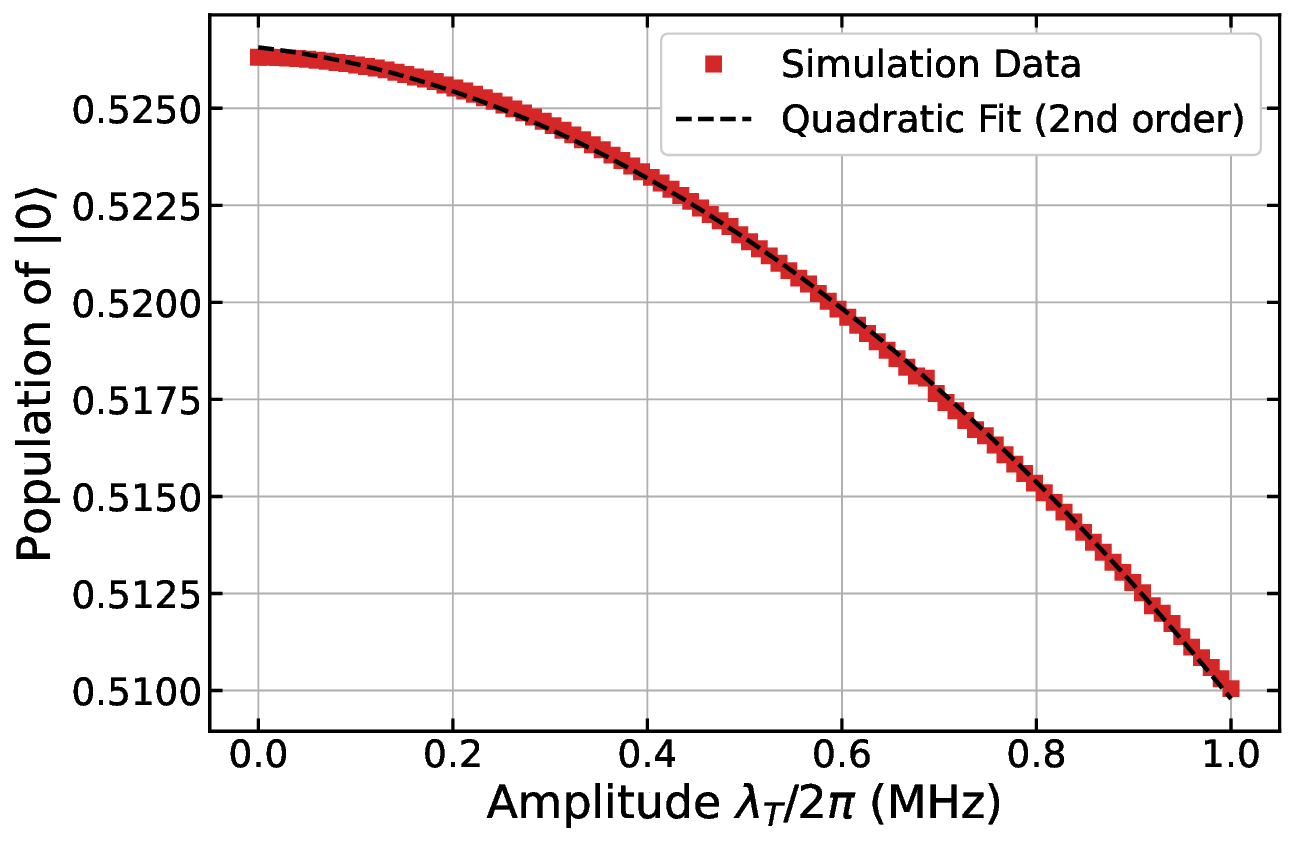}
\caption{Plot of the dependence of the signal intensity on the target-field amplitude $\lambda_T/2\pi$. The parameters are $D/2\pi=2870~\mathrm{MHz}$, $E/2\pi=10~\mathrm{MHz}$, $\lambda_D/2\pi=6.0~\mathrm{MHz}$, $\Gamma/2\pi=2.0~\mathrm{MHz}$, $t=5~\mu\mathrm{s}$, $N_m=T/t$, and $T=1~\mathrm{s}$.}
\label{fig:amplitude_dep}
\end{figure}

Next, we examined how the signal of this sensing scheme responds to the strength of the magnetic field. With the target frequency fixed at the resonance point ($f_T = (2E + \lambda_D/2)/2\pi$), Figure~\ref{fig:amplitude_dep} shows the variation of the signal intensity $P_0$ when the field amplitude ($\lambda_T$) is varied. In the figure, the red squares are simulation results and the black dashed curve is a fit by a quadratic function. As a result, the variation of the signal intensity follows the square of the field amplitude very well. According to Fermi's golden rule, the transition probability from one state to another is proportional to the squared magnitude of the matrix element of the perturbation (the target field). Therefore, the quadratic dependence obtained from the numerical calculations is qualitatively consistent with the analytical result.
 
\subsection{Sensitivity Evaluation under Ideal Projective Measurement}

\begin{figure}[ht]
\centering
\includegraphics[width=0.95\linewidth]{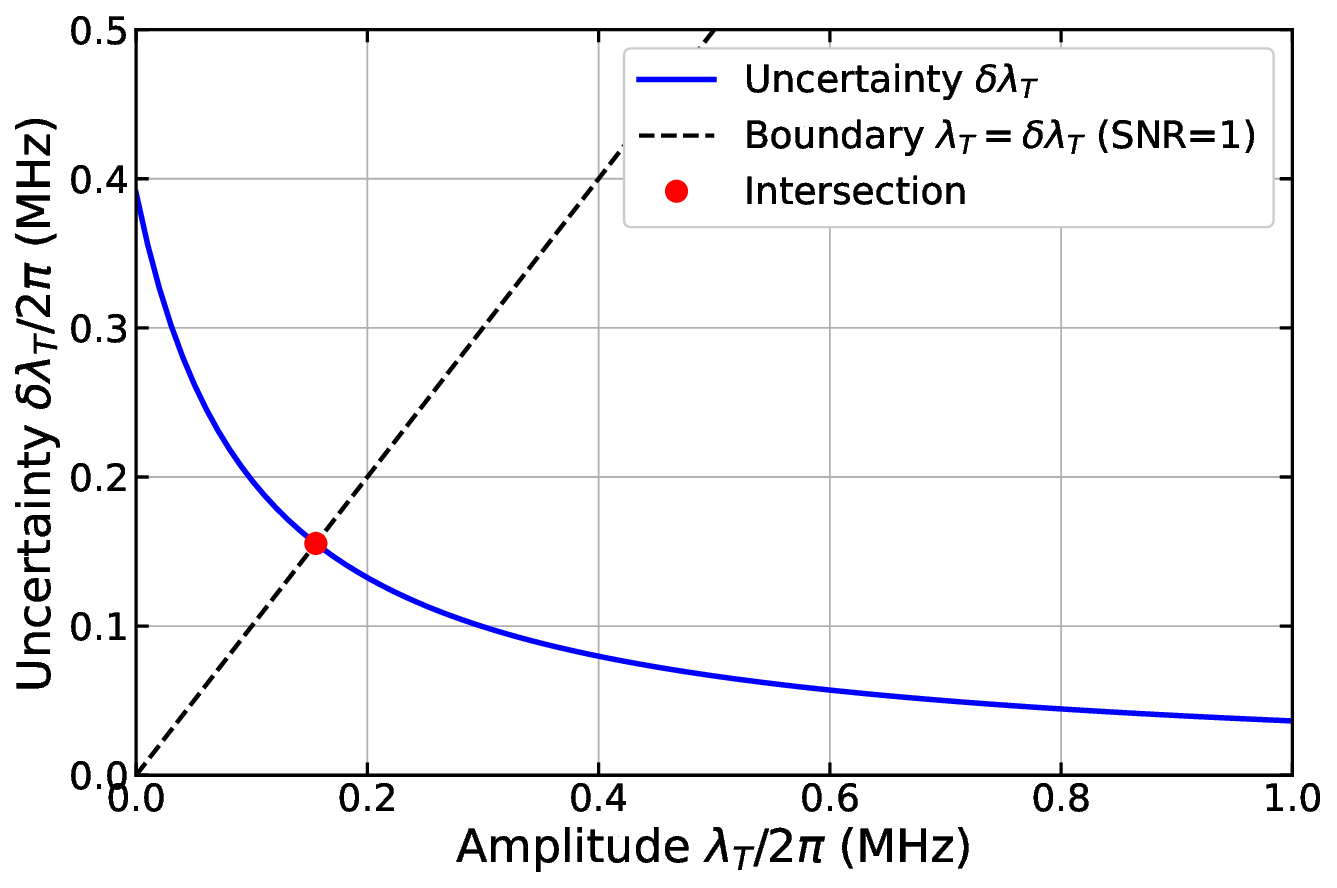}
\caption{Plot of the estimation error as a function of target-field amplitude (solid line). For reference, $\delta\lambda_T = \lambda_T$ is also plotted (dashed line). The parameters are $D/2\pi=2870~\mathrm{MHz}$, $E/2\pi=10~\mathrm{MHz}$, $\lambda_D/2\pi=6.0~\mathrm{MHz}$, $\Gamma/2\pi=2.0~\mathrm{MHz}$, $t=5~\mu\mathrm{s}$, $N_m=T/t$, and $T=1~\mathrm{s}$.}
\label{fig:sensitivity_ideal}
\end{figure}

To quantitatively characterize the performance of our scheme, we first evaluate the sensitivity assuming ideal projective measurement of the spin state. The measurement error is calculated using Eq.~\eqref{eq:dB_ideal}. Here we define the sensitivity (the minimum detectable magnetic field) as the value of $\lambda_T$ at which the estimation error $\delta\lambda_T$ equals the applied field $\lambda_T$. Figure~\ref{fig:sensitivity_ideal} shows the calculated measurement error ($\delta\lambda_T$) as a function of the field amplitude ($\lambda_T$). In the figure, the solid blue line represents the estimation error $\delta\lambda_T$; smaller values mean that smaller variations can be detected and that performance is higher. The graph shows that as the amplitude $\lambda_T$ increases, the value of $\delta\lambda_T$ decreases, corresponding to improved performance. This can be understood from the quadratic dependence of the signal shown in Figure~\ref{fig:amplitude_dep}: when the signal $S$ varies as the square of the amplitude ($S\propto \lambda^2$), its slope is $dS/d\lambda \propto \lambda$, which grows in proportion to the amplitude. A larger slope means that a small change in magnetic field produces a larger change in signal, so the estimation error decreases. The dashed line indicates $\lambda_T = \delta\lambda_T$ (the SNR\,$=1$ boundary), and the region in which the blue line lies below this dashed line is the region of magnetic fields that can be effectively sensed.
 
The intersection point shown by the red dot has a value of $\lambda_T/2\pi = 0.156~\mathrm{MHz}$, from which the sensitivity (the minimum detectable magnetic field for an integration time of one second) is calculated as $5.6~\mu\mathrm{T}/\sqrt{\mathrm{Hz}}$.
 
\subsection{Sensitivity Evaluation Including Measurement Noise and Microwave-Axis Misalignment}

Here we evaluated the sensitivity taking into account two practical imperfections: (i) the finite optical readout efficiency of the spin state, and (ii) a small misalignment of the control-microwave application axis relative to the assumed direction.

\begin{figure}[ht]
\centering
\includegraphics[width=0.95\linewidth]{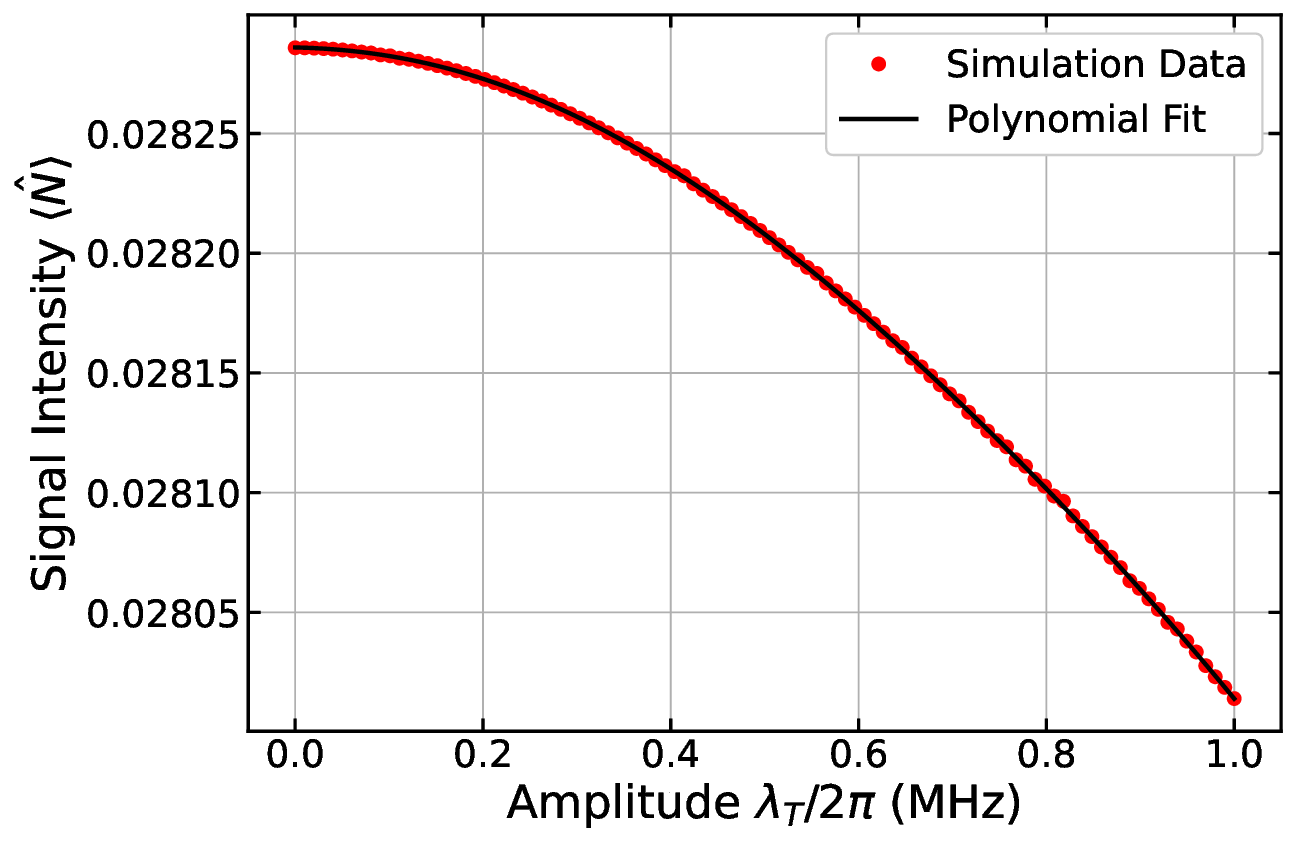}
\caption{Dependence of the signal intensity (expected photon number $\langle\hat{N}\rangle$) described by Eq.~\eqref{eq:N_expectation} on the target-field amplitude $\lambda_T/2\pi$ when measurement noise is taken into account. The parameters are $D/2\pi=2870~\mathrm{MHz}$, $E/2\pi=10~\mathrm{MHz}$, $\lambda_D/2\pi=6.0~\mathrm{MHz}$, $\Gamma/2\pi=2.0~\mathrm{MHz}$, $t=5~\mu\mathrm{s}$, $N_m=T/t$, and $T=1~\mathrm{s}$.}
\label{fig:photon_signal}
\end{figure}

\begin{figure}[ht]
\centering
\includegraphics[width=0.95\linewidth]{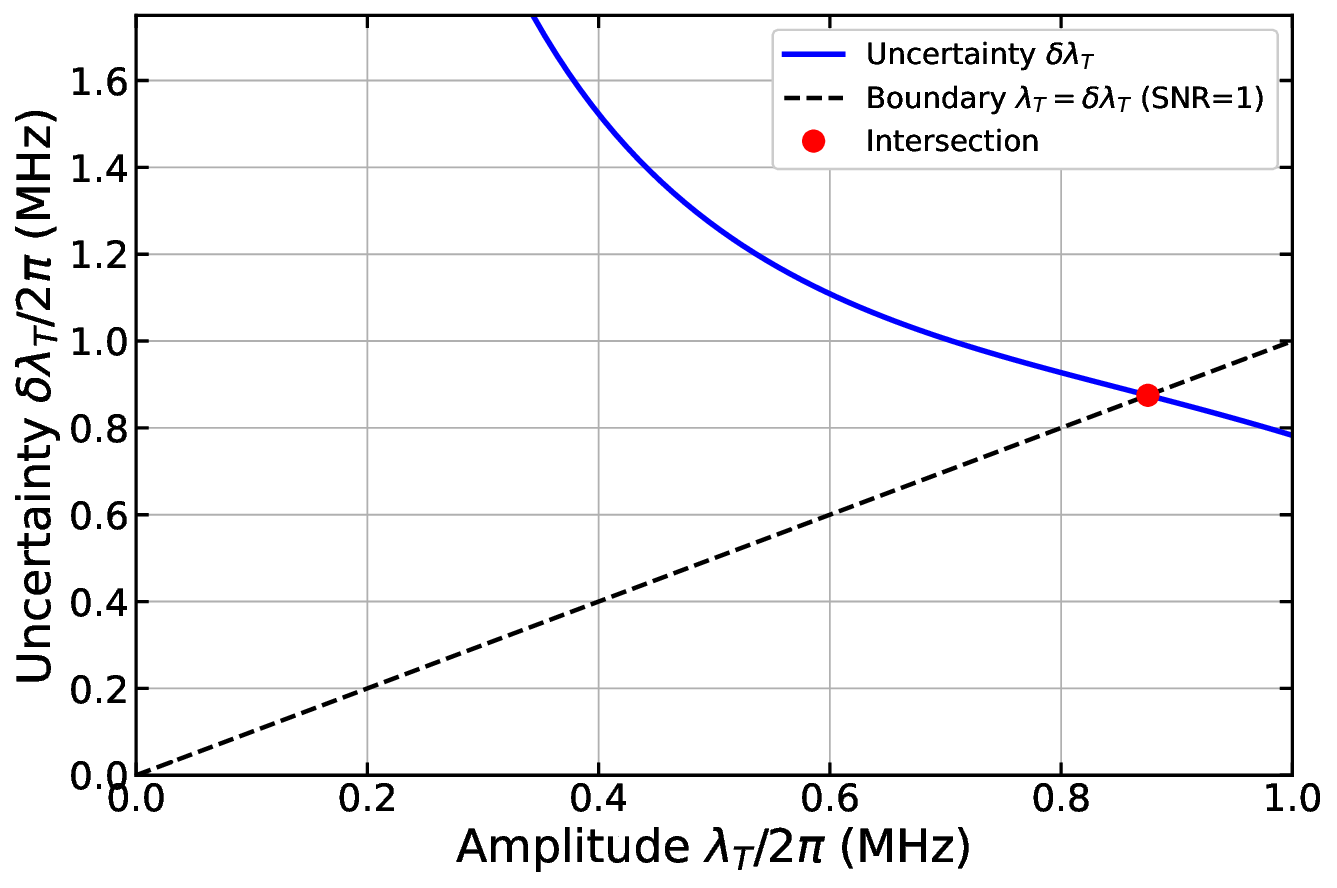}
\caption{Plot of the estimation error as a function of target-field amplitude when measurement noise is taken into account (solid line). For reference, $\delta\lambda_T = \lambda_T$ is also plotted (dashed line). The parameters are $D/2\pi=2870~\mathrm{MHz}$, $E/2\pi=10~\mathrm{MHz}$, $\lambda_D/2\pi=6.0~\mathrm{MHz}$, $\Gamma/2\pi=2.0~\mathrm{MHz}$, $t=5~\mu\mathrm{s}$, $N_m=T/t$, and $T=1~\mathrm{s}$.}
\label{fig:sensitivity_noise}
\end{figure}

\textbf{Effect of measurement noise.}\quad The estimation error is now calculated using Eq.~\eqref{eq:dB_noise}, which accounts for the imperfect optical readout. Figure~\ref{fig:photon_signal} shows the signal intensity (the expected photon number $\langle\hat{N}\rangle$) as a function of the field amplitude, and Figure~\ref{fig:sensitivity_noise} shows the calculated measurement error ($\delta\lambda_T$) as a function of the field amplitude ($\lambda_T$). The estimation error is larger than in the case of ideal projective measurement. In this case, the minimum detectable magnetic field for one-second integration is approximately $30~\mu\mathrm{T}/\sqrt{\mathrm{Hz}}$.
 
\textbf{Effect of microwave-axis misalignment.}\quad In a realistic experimental setup, the control-microwave application direction may deviate slightly from the assumed $y$-axis. To model this, we redefine the Hamiltonian by introducing a misalignment angle $\theta$:
\begin{align}
\hat{H} &= (D+E)\ket{B}\bra{B} + \frac{\lambda_D}{2}\bigl(\cos\theta\,\hat{S}_y + \sin\theta\,\hat{S}_x\bigr) \nonumber\\
&\quad + \frac{\lambda_D}{2}\bigl(-i e^{2i\omega_D t}\ket{D}\bra{0} + i e^{-2i\omega_D t}\ket{0}\bra{D}\bigr) \nonumber\\
&\quad + \lambda_T\bigl(\ket{B}\bra{D}e^{-i\omega_D t} + \ket{D}\bra{B}e^{i\omega_D t}\bigr)\cos\omega_T t.
\label{eq:H_misalign}
\end{align}
For $\theta = 0$, Eq.~\eqref{eq:H_misalign} reduces to Eq.~\eqref{eq:H_sim}. The $\sin\theta\,\hat{S}_x$ term introduces an additional coupling between $\ket{0}$ and $\ket{B}$, which can produce an AC Stark shift of order $(\lambda_D \sin\theta)^2/(8E)$. For the parameter range considered here, this shift is small enough that the resonance condition derived in Chapter~\ref{ch:theory} remains approximately valid.

\begin{figure}[ht]
\centering
\includegraphics[width=0.95\linewidth]{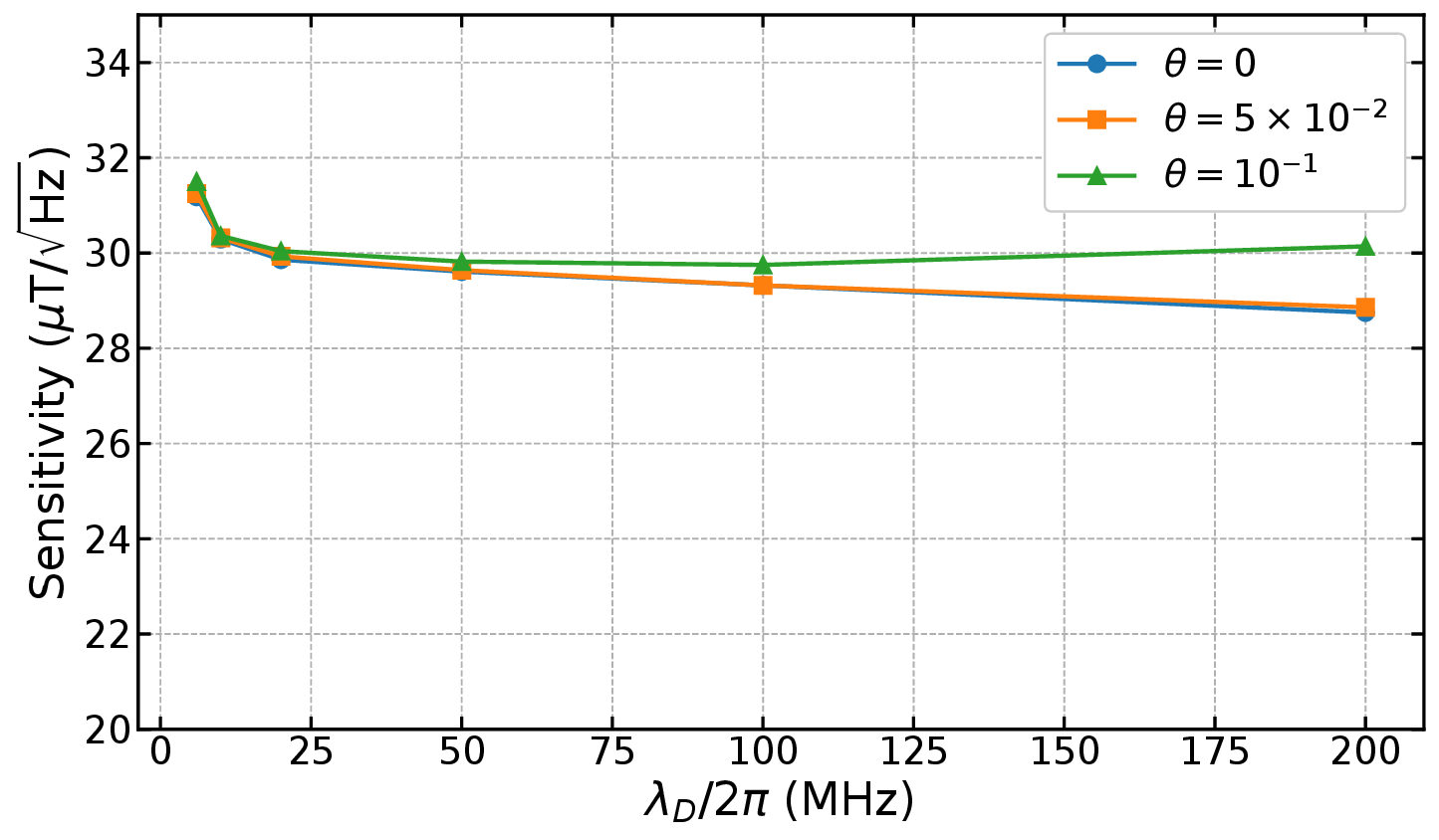}
\caption{Estimated sensitivity as a function of $\lambda_D/2\pi$ for three values of the misalignment angle $\theta = 0$, $5\times 10^{-2}$, and $10^{-1}$~rad, including measurement noise. The other parameters are $D/2\pi=2870~\mathrm{MHz}$, $E/2\pi=10~\mathrm{MHz}$, $\Gamma/2\pi=2.0~\mathrm{MHz}$, $t=5~\mu\mathrm{s}$, $N_m=T/t$, and $T=1~\mathrm{s}$.}
\label{fig:sensitivity_lambdaD}
\end{figure}

Figure~\ref{fig:sensitivity_lambdaD} shows the change in sensitivity as a function of the control-microwave strength $\lambda_D/2\pi$ for $\theta = 0,\, 5\times10^{-2},\, 10^{-1}$~rad, taking the optical readout into account. From the simulation results, even when $\lambda_D/2\pi$ is increased to about $200~\mathrm{MHz}$, the sensitivity remains nearly constant at about $30~\mu\mathrm{T}/\sqrt{\mathrm{Hz}}$, with no significant degradation observed. Moreover, even when $\theta$ is misaligned by about $10^{-1}$~rad ($\approx 5.7^\circ$), almost the same sensitivity is maintained. This result shows that the proposed scheme operates stably under the strong microwave driving needed for broadband operation, and has sufficient tolerance against the small misalignment between the axis and the microwave application direction that is unavoidable in actual experimental environments.
 
\section{Conclusion}

In summary, we have proposed a broadband AC magnetic-field sensing scheme 
based on CW-ODMR that uses microwave-driven dressed states of diamond NV 
centers. In previous studies, the bandwidth of AC magnetic-field sensors based on CW-ODMR was limited to about $10~\mathrm{MHz}$. In contrast, our scheme uses microwave driving, in which the Rabi frequency can in principle be increased to about $100~\mathrm{MHz}$, so that not only AC magnetic fields with frequencies of a few MHz but also AC magnetic fields with frequencies of the order of $100~\mathrm{MHz}$, which has been difficult to achieve previously, can be detected. Numerical calculations were used to evaluate this performance quantitatively, and we found that, even when measurement noise is taken into account, a sensitivity of $30~\mu\mathrm{T}/\sqrt{\mathrm{Hz}}$ per single NV center are possible. This scheme is expected to find applications in a wide range of fields, including biological applications and materials engineering.

This project is supported by
    JST Moonshot R\&D Grant
    Number JPMJMS226C, 
    JST CREST Grant Number JPMJCR23I5, Presto
    JST Grant Number JPMJPR245B, and
JSPS Grant-in-Aid for JSPS Fellows (Grant No. JP24KJ1964).
This word is partly supported by MEXT Q-LEAP (Grant No. JPMXS0118067395), JSPS KAKENHI (Grant No. JP25K01292), JSPS Grant-in-Aid for JSPS Fellows (Grant No. JP24KJ1964), JST CREST (Grant No. JPMJCR24A5), JST ASPIRE (Grant No. JPMJAP2427), the Cross-ministerial Strategic Innovation Promotion Program (SIP) Program, Center for Spintronics Research Network (CSRN), Keio University and the Program for the Advancement of Next Generation Research Projects, Keio University.
\bibliography{ref}

\end{document}